\documentclass[12pt]{article}
\usepackage{cite}
\usepackage{graphicx}
\usepackage{ amssymb } 
\pdfoutput=1

\begin{document}

\newcommand{\titulo}{A scenario for the critical fluctuations\\ near the transition of few-bilayer  films\\ of high-temperature cuprate superconductors}

\newcommand{\autor}{M.M.~Botana $^{a,b}$, M.V.~Ramallo$^{a,b}$}

\newcommand{\direccion}{$^{a}$Quantum Materials and Photonics Research Group (QMatterPhotonics),\\ Department of Particle Physics, University of Santiago de Compostela,\\ 15782 Santiago de Compostela, Spain \\ \mbox{}\\
$^{b}$Instituto de Materiais (iMATUS),\\ University of Santiago de Compostela, 15782 Santiago de Compostela, Spain}

\begin{center}
  \Large\bf
\titulo\\  \end{center}\mbox{}\vspace{-1cm}\\ 

\begin{center}\normalsize\autor\end{center} 

\begin{center}\normalsize\it\direccion\end{center}

\newcommand{\abs}[1]{\left\vert#1\right\vert} 
\newcommand{\parent}[1]{\left( #1 \right)}
\newcommand{\corch}[1]{\left[ #1 \right]}
\newcommand{\llave}[1]{\left\{ #1 \right\}}
\newcommand{\unit}[1]{{\rm  } \mathrm{#1}}
\newcommand{\function}[2]{#1 \left( #2 \right)}
\newcommand{\prom}[1]{\langle#1\rangle}

\newcommand{\dd}{\mbox{\rm{d}}}
\newcommand{\DF}{\mbox{$\Delta F$}}
\newcommand{\xx}{\mbox{${\rm x}$}}
\newcommand{\yy}{\mbox{${\rm y}$}}
\newcommand{\xy}{\mbox{${\rm xy}$}}
\newcommand{\Tc}{\mbox{$T_{{\rm mf}}$}}
\newcommand{\Tkt}{\mbox{$T_{{\rm KT}}$}}
\newcommand{\Df}{\mbox{$\Omega$}}

\newcommand{\Ds}{\mbox{$\Delta\sigma$}}
\newcommand{\DsGGL}{\mbox{$\Delta\sigma_{\rm GGL}$}}
\newcommand{\DsKT}{\mbox{$\Delta\sigma_{\rm KT}$}}

\newcommand{\Dspure}{\mbox{$\Delta\sigma_{\rm hom}$}}
\newcommand{\Tcprime}{\mbox{$T'_{{\rm mf}}$}}

\newcommand{\electron}{\mbox{e}}

\newcommand{\Gi}{{\rm Gi}}

\newcommand{\DT}{\mbox{$\Delta_{\rm KT}$}}
\newcommand{\ev}{\mbox{$\omega$}}

\newcommand{\Nuc}{{\cal N}} 
\newcommand{\Nuctext}{\mbox{$\Nuc$}}

\newcommand{\cc}{\mbox{$\varepsilon^c$}}
\newcommand{\ybco}{\mbox{YBa$_2$Cu$_3$O$_{7-\delta}$}}
\newcommand{\pbco}{\mbox{PrBa$_2$Cu$_3$O$_{7-\delta}$}}
\newcommand{\bscco}{\mbox{Bi$_2$Sr$_2$CaCu$_2$O$_{8+x}$}}
\newcommand{\cuodos}{\mbox{CuO$_2$}}

\newcommand{\eps}{\mbox{$\varepsilon$}}

\newcommand{\gamU}{\mbox{$\gamma_{\rm int}$}}
\newcommand{\gamD}{\mbox{$\gamma_{\rm ext}$}}
\newcommand{\gamUcuad}{\mbox{$\gamma_{\rm int}^2$}}
\newcommand{\gamDcuad}{\mbox{$\gamma_{\rm ext}^2$}}
\newcommand{\gamUcubo}{\mbox{$\gamma_{\rm int}^3$}}
\newcommand{\gamDcubo}{\mbox{$\gamma_{\rm ext}^3$}}
\newcommand{\gamUcuar}{\mbox{$\gamma_{\rm int}^4$}}
\newcommand{\gamDcuar}{\mbox{$\gamma_{\rm ext}^4$}}

\newcommand{\cfl}{\mbox{$c_{\rm\, fl}$}}
\newcommand{\chifl}{\mbox{$\chi_{\rm\, fl}$}}
\newcommand{\sigmafl}{\mbox{$\sigma_{\rm\, fl  AL}$}}

\newcommand{\eg}{{e.g.}}
\newcommand{\ie}{{i.e.}}
\newcommand{\lsim}{\mbox{$\stackrel{<}{_\sim}$}}
\newcommand{\gsim}{\mbox{$\stackrel{>}{_\sim}$}}
\newcommand{\etal}{{et~al}}

\newcommand{\proptosim}{\mathrel{\vcenter{\offinterlineskip\halign{\hfil$##$\cr\propto\cr\noalign{\kern1pt}\sim\cr\noalign{\kern-1pt}}}}} 


\mbox{}\vskip0.5cm{\bf Abstract: }

We study the critical fluctuations near the resistive transition  of very thin films of high-temperature cuprate superconductors composed of a number \Nuctext\ of only a few unit cells of superconducting bilayers. For that, we solve the fluctuation spectrum of a Gaussian--Ginzburg--Landau model for few-bilayers superconductors considering two alternating Josephson interlayer interaction strengths, and we obtain the corresponding paraconductivity above the transition. Then, we extend these calculations to temperatures below the transition through expressions for the Ginzburg number and Kosterlitz--Thouless-like critical region. When compared with previously available data in \ybco\ few-bilayers systems, with \Nuctext\,= 1 to 4, our results  seem to  provide a plausible scenario for their critical regime.

\vspace{4cm}

\mbox{}\hfill{\footnotesize {\tt mv.ramallo@usc.es}}
\thispagestyle{empty}

\newpage
\setlength{\baselineskip}{18pt}
\




\section{Introduction\label{sec:introduccion}}


The study of critical fluctuations near the transition temperature in  {high-temperature cuprate superconductors}, HTSC, has attracted much interest since the discovery of these materials~\cite{N12,N3,N16,A13,B1,B2,B3,B4}. In HTSC, these critical effects are especially significant  due, mainly, to~the short coherence lengths and corresponding reduced-dimensionality enhancements when competing with the size of the intrinsic layered nanostructure formed by the \cuodos\ superconducting planes~\cite{N12,N3,N16,A13,klemm,buzdin,R25}. It was quite  early noted that the temperature behavior of the critical fluctuations (including both critical exponents and amplitudes) could provide  information about HTSC such as, \eg, the~locus where superconductivity occurs, the~symmetry of the pairing wave function, or~the possible influence of phase fluctuations on the high value of transition temperature itself~\cite{N12,N3,N16,A13,R25,A11,A19,klemm,buzdin,B1,B2,B3,B4}.  Today, theories and corresponding equations are available that quite satisfactorily account  for  the roundings near the transition of key observables, such as the electrical resistivity, in~regular bulk HTSC samples, \ie, those with a macroscopic number of superconducting planes (see, \eg,~\cite{R25,A11,A19,klemm}).

\begin{figure}[h!]
\includegraphics[width = 0.95\textwidth]{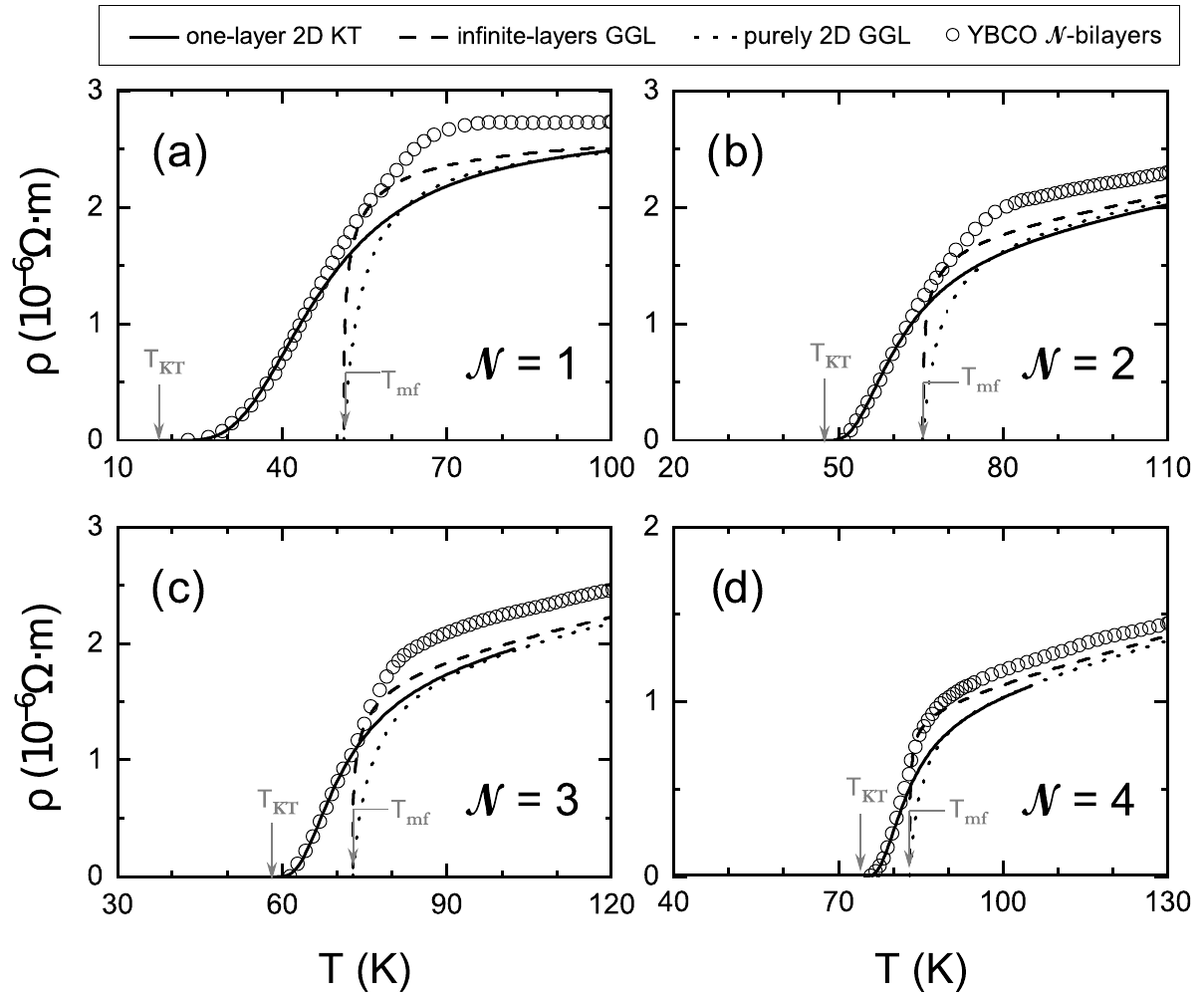}
\caption{ {Electrical resistivity} 
 $\rho$ vs.~$T$ obtained experimentally by Cieplak \etal.~\cite{cieplak} in samples  {with {\bf (a)} \Nuctext=1,  {\bf (b)} \Nuctext=2,  {\bf (c)} \Nuctext=3 and  {\bf (d)} \Nuctext=4}  unit cells of superconducting bilayers of  \ybco\  (open circles;  {taken} from {Figure~6b}  of~\cite{cieplak}). The~solid line is a fit using the classical one-layer 2D  prediction for Kosterlitz--Thouless (KT) critical fluctuations in the tail of the transition~\cite{HN}. The agreement is excellent in the lower part of the transition, although~with a large variation of the  amplitude parameter $A_{\rm KT}\simeq950, 6500, 6500, 9000 \,(\Omega {\rm m})^{-1}$ for \Nuctext\,= 1 to 4, respectively. The~dot--dashed line is a fit using the conventional Lawrence--Doniach prediction for the Gaussian--Ginzburg--Landau (GGL)   fluctuations of  superconductors composed of a macroscopic number of layers~\cite{LD,R25}. In contrast with what happens in thick films or crystals of \ybco, the~infinite-layers GGL prediction is only a tangent to the data. Lowering its fitting-region temperatures to more smoothly connect with the KT results would only worsen the quality of the overall fit. Imposing a fully 2D behavior also worsens the fit (the dotted line corresponds to $\xi_c(0) = 0$ in the Lawrence--Doniach result). This comparison suggests that  considering a finite number of layers in the theory predictions will be needed to fully account for the GGL region (and also to justify the $A_{\rm KT}$ variation). See Section~\ref{sec:introduccion} for a description of the equations and free parameters used in the fits in this figure. See Figure~\ref{fig:fitted_R} for the fits to the same data  with  the  expressions obtained in this paper for few-bilayers superconductors.
\label{fig:intro}
}
\end{figure}

However, the~understanding of the critical superconducting effects in very thin films of HTSC, composed of a number \Nuctext\ of only a few ($\Nuc\,\lsim\,5$)  unit cell layers of the material,  is much less  established. Those few-layers HTSC are today growable by a number of different techniques (usually either built on a substrate or sandwiched into heterostructures, or~also obtained via surface gating) \cite{R13,R17,R4,cieplak,R6,R7,R8,R9,R11,R3,R35,pavuna}. Experimentalists measuring the resistive transition of their few-layers HTSC have up to now focused mainly on identifying the most unambiguous feature of two-dimensionality (2D) in their samples, which happens in the $T$-region corresponding to the  $\rho\rightarrow0$ tail in the electrical in-plane resistivity-versus-temperature curves, $\rho(T)$.  That region becomes wider and displays a characteristic  exp-like divergence of the electrical conductivity, which is a landmark feature of the enlargement of the transition  due to vortex--antivortex interactions famously predicted by \mbox{Berezinskii~\cite{berezinskii1,berezinskii2}}, Kosterlitz and Thouless~\cite{KT} (KT) for 2D complex order parameters (and then for superconductors by, \eg,~\cite{R22, HN}). However,~apart from this success with the transition tail, the~understanding of the whole $\rho(T)$ transition is today still somewhat  lacking. Let us now, for~introductory purposes,  briefly  comment on what we believe are the  main currently open issues, for~which we will use the help of our Figure~\ref{fig:intro}.

In Figure~\ref{fig:intro}, we represent the $\rho(T)$ data obtained in the pioneering work of Cieplak \etal.~\cite{cieplak} in samples comprising  \Nuctext\,=  1 to  4 unit cells of the prototypical HTSC compound  \ybco\ (YBCO). Note that  every unit cell of YBCO comprises two \cuodos\ superconducting  layers~\cite{leggett}. We also plot (solid lines) the best fit to the tail $\rho\rightarrow0$ of the transition using the classical KT equation~\cite{HN} $\rho^{-1}  =  \rho^{-1}_n + A_{\rm KT} \exp\sqrt{\beta/(T-\Tkt)}$, being $A$, $\beta$ and \Tkt\  free parameters (and $\rho_n$ the normal-state resitivity, \ie, the~one without critical fluctuations, that in these samples  is easy to obtain~\cite{cieplak}  as a linear extrapolation of the behavior of $\rho$ at higher temperatures). As~was indeed already noticed in~\cite{cieplak}, this produces an excellent agreement with the data in the lower part of the transition. In addition, the~so-obtained KT transition temperature \Tkt\ is  in good agreement with the temperature at which the signal ceases to be ohmic, which is another distinguishing feature of the KT transition~\mbox{\cite{HN,R21,R35,cieplak}.} All of this indicates that the samples are thin enough to display some 2D-like~behavior. 

To our knowledge, it remains to be explained  why this agreement is obtained only assuming a very large variation of  the  KT amplitude $A_{\rm KT}$ with \Nuctext\ (about one order of magnitude  from  \Nuctext\,= 1 to \Nuctext\,=  4, see values in the caption of Figure~\ref{fig:intro}).

However, even more important (and as already indicated by Cieplak \etal\ themselves~\cite{cieplak}), the roundings of the mid-to-upper part of the transitions do not adhere to the KT behavior. Thus, for~those temperatures, an~explanation in terms of different fluctuation theories, such as the Gaussian--Ginzburg--Landau (GGL) approach, seems to be necessary. 
{In that approach, small excitations of the order parameter are considered into the GL expressions of the thermal averages, as~described in detail, \eg, in~\cite{buzdin,klemm,R25,A19,LD}  (or into microscopic diagramatic approaches~\cite{ALarkin,HLarkin,MakiT,MThompson} with equivalent results, especially for non-$s$-wave pairing where anomalous Maki--Thompson contributions become negligible~\cite{R25,MakiT,MThompson,Yip,A19}).} 
However, the~existing GGL equations do not seem to fit these data (in contrast to their success in bulk HTSC~\cite{R25,A11,A19,klemm}). This is also shown in our Figure~\ref{fig:intro}: There, we use the equation due to Lawrence and Doniach~\cite{LD} for the GGL fluctuation-induced conductivity in layered superconductors of macroscopic size (\ie, infinite-layers  superconductors), namely $\rho^{-1}  =  \rho^{-1}_n + {\rm e}^2/(16\hbar d\sqrt{\varepsilon^2+B\varepsilon})$, where  $d$ is the average interlayer distance (5.85 \AA\ in YBCO), $\varepsilon = \ln(T/\Tc)$, \Tc\ is a mean-field critical temperature  and $B\equiv(2\xi_c(0)/d)^2$ is  a  constant that involves  the inter-plane coherence length amplitude $\xi_c(0)$ ({{in all of this paper, e,}  $\hbar$ and $k_{\rm B}$ are the usual physical constants}). As~illustrated by Figure~\ref{fig:intro} (dot--dashed line), the~equation fails to continue the good fit achieved  by the KT approach. (Note that, in~contrast, this GGL equation does succeed in fitting this transition region  in bulk, infinite-layers YBCO with $\xi_c(0)\simeq1$ \AA, as~shown by various authors~\cite{R25,A11,A19,klemm}.) Imposing in that GGL result a 2D condition is possible by imposing $\xi_c(0) = 0$, but~this also does not improve the GGL fit, as~ shown as well in Figure~\ref{fig:intro}  {(dotted line).} {{The failure of the GGL approach for infinite-layers superconductors when applied to} finite-layers samples was in fact already noted by Cieplak \etal.~\cite{cieplak} (they also explored to solve these discrepancies by testing whether critical-temperature inhomogeneities could explain them, but~they demonstrated instead that a random spatial distribution of such inhomogeneities could not account for the differences; only a handpicked, difficult to justify spatially ordered distribution of inhomogeneities {{in series} } could make the infinite-layers theory agree with the data).}

It seems evident, therefore, that  to understand the whole resistive transition of few-layers YBCO, it is necessary to develop a GGL calculation explicitly taking into account the finiteness of their number of superconducting planes. The~purpose of the present paper is to present that theoretical development and compare it with available data, so to propose what is, we believe, a~rather plausible scenario for the resistive transition rounding in these~systems.

Let us also note here that a first, but~incomplete, attempt was presented by some of us in a past Conference Proceeding~\cite{alberto} in which we solved the GGL fluctuation spectrum for a limited set of few-layers cases. However, our  conclusion there was  that the calculation would be feasible in full only up to the three-layers case (thus only up to \Nuctext\,= 1 for YBCO). In~contrast, in~the present paper, we will show that by  focusing on interlayer Josephson coupling strengths that take two alternating values (the case expected for YBCO, and~in fact for all  HTSC with two \cuodos\ layers per unit cell~\cite{leggett,R25,klemm,buzdin}), it is possible to obtain explicit expressions for a much larger, and~useful, number of layers. Additionally,  we will consider an extension of these results to the important KT regime (to also explain the lower temperature region of the transition) and the inclusion of an energy cutoff (to also obtain agreement at higher temperatures).

The organization of the present paper is as follows.  Section~\ref{sec:teoria}  is devoted to our theory calculations: in particular, in~{Section}  \ref{ssec:spectrum}, we present our starting GGL model for few-bilayers HTSC and  calculate its spectrum of fluctuations; then, in {Section}~\ref{ssec:DsGGL}, we calculate the resulting GGL fluctuation electrical conductivity; in {Section}~\ref{ssec:Gi}, we consider the important aspect of the temperature of crossover toward non-GGL KT-like fluctuations   (\ie, the~Ginzburg number) and its dependence on the number of bilayers \Nuctext; in {Section}~\ref{ssec:DsKT}, we extend these results to the KT region of the fluctuations, obtaining expressions that explicitly take into account the few-bilayers effects and predict values for the effective KT amplitudes of the fluctuation conductivity; and in {Section}~\ref{ssec:EMA}, for completeness, we discuss the effects of possible critical-temperature inhomogeneities on these theory results. Then, in~Section~\ref{sec:experimental}, we  compare these theory developments with an example of experimental data of the resistive transition of few-bilayers {YBCO}, for~which we use the paradigmatic data of Cieplak \etal.~\cite{cieplak}. {(In addition, in~an~Appendix, we  compare our equations with data available~\cite{bscco_r1} for  few-bilayers \bscco\ (BSCCO).)} Finally, in~Section~\ref{sec:conclusiones}, we summarize some conclusions, {implications and possible further research}  suggested by our~results.


\section{Calculation of the  Fluctuation Electrical Conductivity of a HTSC Composed of \boldmath\Nuctext-Bilayers in the Gaussian--Ginzburg--Landau GGL and Kosterlitz--Thouless KT-like~Regimes\label{sec:teoria}}



\subsection{Spectrum of Fluctuations above the Mean-Field Critical Temperature \Tc\ \/ for Few-Bilayers Superconductors in a Gaussian--Ginzburg--Landau (GGL) Approximation\label{ssec:spectrum}}

We take as the starting point of our modelization a Ginzburg--Landau (GL) free energy functional that considers a finite number (\Nuctext) of layered unit cells of an HTSC having two superconducting layers per unit cell (such as YBCO, where each layer corresponds to a \cuodos\ plane). We label each of those layers with a double index $jn$, where\linebreak $n = 1\dots\Nuc$ indicates the unit cell and $j = 1,2$ signals the layer inside the cell. We associate a superconducting wave function $\psi_{jn}$ to each layer. For~the interlayer interactions, we adopt the same common Josephson-type coupling as the usual Lawrence--Doniach model for infinite-layers systems, but~considering different intra-cell and inter-cell coupling strength constants, \gamU\ and \gamD.  The~corresponding GL functional, in~the Gaussian approximation above its transition temperature (henceforth called mean-field critical temperature and noted \Tc\ to better distinguish it from the KT vortex-antivortex temperature \Tkt\ that we shall introduce later), is then:\vspace{-3pt}
\begin{equation}
	\label{eq:DF}
	\DF  =  \sum_{n = 1}^{\Nuc} \sum_{j = 1}^{2}  \DF_{jn}^{\rm 2D} 
+ \sum_{n = 1}^{\Nuc}   \DF_n^{\rm int} 
+ \sum_{n = 1}^{\Nuc-1}   \DF_n^{\rm ext}.
\end{equation}
where $\DF_{jn}^{\rm 2D}$, $\DF_n^{\rm int}$ and $\DF_n^{\rm ext}$ are contributions due to, respectively, the~in-plane interactions, intra-cell interlayer interactions, and~extra-cell interlayer interactions:\vspace{-3pt}
\begin{equation}
	\label{eq:DFintr}
	\DF_{jn}^{\rm 2D}  =  a_0\int \dd^2 \mbox{\bf r}\;  \left\{ \varepsilon |\psi_{jn}|^2 +  \xi^2_{ab}(0)  |\nabla_{xy} \psi_{jn}|^2 \right\},
\end{equation}
\begin{equation}
\DF_n^{\rm int}  =  a_0\int \dd^2 \mbox{\bf r}\; \gamU | \psi_{2n} - \psi_{1n} |^2,
\end{equation}
\begin{equation}
\DF_n^{\rm ext}  =  a_0 \int \dd^2 \mbox{\bf r}\; \gamD | \psi_{1,n+1} - \psi_{2n} |^2.
\end{equation}
{In these}  equations, ${\bf r}$ is the in-plane coordinate, $\nabla_{xy}$ the in-plane gradient, $\xi_{ab}(0)$ is the GL amplitude of the in-plane coherence length, $a_0$ is the GL normalization constant and $\varepsilon$ is the reduced temperature that we take as 
\begin{equation}
\varepsilon  =   \ln\, \left(T/\Tc\right).
\end{equation}
{This} choice of $\varepsilon$ is usual when analyzing data that include the $\varepsilon\,\gsim\,0.1$ temperature region well above the transition, as~it usually improves the agreement with the data and is supported by the microscopic derivations of the GL equations. When $\varepsilon\,\lsim\,0.1$, this reduces  to the limit $\varepsilon\approx (T-\Tc)/\Tc$ usually found in many~textbooks.

Obviously, the equilibrium (minimum \DF) given by that functional above \Tc\ is just $\psi_{jn}^0 = 0$ (\ie, fully normal state); to obtain the critical fluctuations, we  must calculate the energy of excitations $\psi_{jn}\neq0$. For~that, we apply the common approach~\cite{LD} of   decomposing them as fluctuation modes additive in energy by~first writing the functional in Fourier space and then diagonalizing the matrix that arises from  the interlayer interaction terms. A~similar approach may be found for other cases of layered geometries in~\cite{R25,klemm,alberto,buzdin}. In~particular, we expand the order parameter through  \mbox{$
\psi_{jn}^\alpha  =  \sum_{\alpha\bf k } \psi_{jn\bf k}^\alpha e^{i\bf k \bf r}$}, where the index $\alpha$ labels the real and imaginary components, and~$\bf k$ is an in-plane wavevector. This leads to
\begin{equation}  
\DF  =  a_0 \sum_{\alpha = Re,Im}  \int \dd^2 \mbox{\bf k}\; \left[ \sum_{j\,n} \,\left(\varepsilon + \xi^2_{ab}(0)k^2\right)\,  | \psi^\alpha_{jn {\rm \bf k}}|^2 +  \sum_{j\,n,j'n'} \,\Df_{j\,n,j'n'} \,\psi_{jn{\rm \bf k}}^{\alpha\mbox{*}}   \psi^{\alpha}_{j'n'{\rm \bf k}}   \; \right],
\label{DFMatrizSinDiagonalizar1}
\end{equation}
where the $\Df_{j\,n,j'n'}$ are  given by the 2\Nuctext $\times$ 2\Nuctext\ matrix 
\begin{equation}
\Df =  
    \left(
    \begin{array}{cccccc}
     \gamU   &   -\gamU          \\
     -\gamU\,\mbox{}   &   \gamU+\gamD  & -\gamD  & & {\rm \large 0} \\       
     &  -\gamD   &   \gamU+\gamD & -\gamU     \\
      & &      -\gamU  &  \ddots     \\
        &   {\rm \large 0} & & &    \gamU+\gamD  & -\gamU   \\ 
     &   & & &    -\gamU  & \gamU   \\
    \end{array}
    \right).
\label{DFMatrizSinDiagonalizar2}
\end{equation}
{Equations}~(\ref{DFMatrizSinDiagonalizar1}) and~(\ref{DFMatrizSinDiagonalizar2}) may be now diagonalized so to obtain the desired expression of the GGL functional in terms of energy-additive fluctuation modes:
\begin{equation} 
\label{eq:DFk}
\DF  =  a_0 \sum_{\alpha jn}  \int \dd^2 {\bf k}\;  \left(\varepsilon + \xi_{ab}^2(0)k^2 + \ev_{jn}\right)\;\left| f_{jn{\bf k}}^\alpha \right|^2 ,
\end{equation}
where $\ev_{jn}$ are the 2\Nuctext\ eigenvalues of the $\Df_{j\,n,j'n'}$ matrix, and~$f_{jn{\bf k}}^\alpha$ is its set of eigenvectors. Obviously, this equation will be useful only as far as the explicit diagonalization of the  $\Df_{j\,n,j'n'}$ matrix is feasible. In~principle, this could be nontrivial for arbitrary \Nuctext, because~it requires finding the zeroes of a polynomial  of degree $2\Nuc$.  However, we found that it is actually possible to carry out the diagonalization for, at~least, \Nuctext  =  1 to 12. The~algebra and the final expressions for $\ev_{jn}$ are unsurprisingly very long, but~software may be used to ease its processing. For~concreteness (and because of the data to be analyzed in the next Sections), we write here the explicit  results  for \Nuctext \,= 1 to 4.

For \Nuctext  =  1:
\begin{eqnarray}
\ev_{\;1\;1} & = & 0 \label{EspectrosInicio}\\
\ev_{\;2\;1} & = & 2 \gamU
\end{eqnarray}

For \Nuctext  =   2:
\begin{eqnarray}
\ev_{\;1\;1} & = & 0 \\
\ev_{\;2\;1} & = & 2 \gamU \\
\ev_{\;1\;2} & = & \gamU+\gamD - \sqrt{\gamUcuad+\gamDcuad} \\
\ev_{\;2\;2} & = & \gamU+\gamD + \sqrt{\gamUcuad+\gamDcuad}
\end{eqnarray}

For \Nuctext  =   3:
\begin{eqnarray}
\ev_{\;1\;1} & = & 0 \\
\ev_{\;2\;1} & = & 2 \gamU \\
\ev_{\;1\;2} & = & \gamU+\gamD-\sqrt{\gamUcuad-\gamU \gamD+\gamDcuad} \\
\ev_{\;2\;2} & = & \gamU+\gamD+\sqrt{\gamUcuad-\gamU \gamD+\gamDcuad} \\
\ev_{\;1\;3} & = & \gamU+\gamD-\sqrt{\gamUcuad+\gamU \gamD+\gamDcuad} \\
\ev_{\;2\;3} & = & \gamU+\gamD+\sqrt{\gamUcuad+\gamU \gamD+\gamDcuad}
\end{eqnarray}

For \Nuctext  =  4:
\begin{eqnarray}
\ev_{\;1\;1} & = & 0 \\
\ev_{\;2\;1} & = & 2 \gamU \\
\ev_{\;1\;2} & = & \gamU+\gamD-\sqrt{\gamUcuad+\gamDcuad} \\
\ev_{\;2\;2} & = & \gamU+\gamD+\sqrt{\gamUcuad+\gamDcuad} \\
\ev_{\;1\;3} & = & \gamU+\gamD-\sqrt{\gamUcuad+\gamDcuad-\sqrt{2} \gamU \gamD} \\
\ev_{\;2\;3} & = & \gamU+\gamD+\sqrt{\gamUcuad+\gamDcuad-\sqrt{2} \gamU \gamD} \\
\ev_{\;1\;4} & = & \gamU+\gamD-\sqrt{\gamUcuad+\gamDcuad+\sqrt{2} \gamU \gamD} \\
\ev_{\;2\;4} & = & \gamU+\gamD+\sqrt{\gamUcuad+\gamDcuad+\sqrt{2} \gamU \gamD}  \label{EspectrosFinal}
\end{eqnarray}
{Let us also note} that in a previous conference-proceedings paper~\cite{alberto}, we presented a similar treatment for few-layers superconductors leading to a similar diagonalization problem that we could solve in full only up to the 3-layers case (thus only up to \Nuctext\,= 1 in the context of this paper). What makes now our present problem explicitly diagonalizable up to, at~least, $\Nuc = 12$ (a 24-layers case) is the alternation of the values \gamU\ and \gamD\ in the matrix of Equation~(\ref{DFMatrizSinDiagonalizar2}). This produces factorizations in the eigenvalues equation making it explicitly solvable.

\mbox{}  

\subsection{Gaussian-Ginzburg-Landau Paraconductivity \DsGGL\label{ssec:DsGGL}}

Once the GGL free energy has been obtained in terms of a fluctuation spectrum of independent fluctuation modes,  it may be possible to calculate fluctuation-induced observables. In~this paper, we focus on the so-called  paraconductivity \Ds, which is defined as~\mbox{\cite{klemm,R25,A11,A19}}
\begin{equation}
\Ds\equiv\rho^{-1}-\rho_n^{-1},
\end{equation}
where $\rho$ is the  the in-plane electrical resistivity and $\rho_n$ is its normal-state background (\ie, the~resistivity that would exist in absence of superconducting effects, that should be obtainable, \eg, by~extrapolating the high-temperature behavior).  From~an experimenter point of view, \Ds\ is one of the most reliable fluctuation-induced observables that may be measured in a few-bilayers HTSC  (note, \eg, that the  heat capacity or  the magnetic moment  are  expected to give very low signals in so tiny samples~\cite{loram,mgb}). The~paraconductivity in bulk HTSC has also been extensively measured and~successfully accounted for in terms of GGL calculations for~temperatures above \Tc\ (see, \eg,~\cite{klemm,R25,A11,A19}). 

 Base formalisms are  well-established to calculate \Ds\ in the GGL approximation in any layered case once their interlayer spectrum is known; in particular, we will use its standard relationship with the summation of the reciprocals of  $\varepsilon+\ev_{jn}$ (see, \eg, {Ref.}~\cite{R25} for a detailed exposition   rewritable with relative ease for the few-bilayers case): 
\begin{equation}
\label{eq:quantity}
\DsGGL  =  \frac{e^2}{32 \hbar d\Nuc} 
\sum_{jn} \left(
\frac{1}{\varepsilon + \ev_{jn}}-
\frac{1}{\cc + \ev_{jn}}
\right).
\end{equation}
{Here,} $2d$ is the thickness of a layered unit cell (\ie, $d$ is the average of the intra-cell and inter-cell interlayer distances). For~the $jn$ summation and $\ev_{jn}$ spectrum, the~results obtained for each \Nuctext\ in the previous subsection are to be used. Note also that for completeness, Equation~(\ref{eq:quantity}) includes a total-energy cutoff \cc\  accounting for the effects of short-wavelength fluctuations, which are expected to be relevant only for temperatures sufficiently above \Tc~\cite{carballeira,A13,A11,A19,mgb}. The corresponding result without a cutoff may be recovered simply as the $\cc\rightarrow\infty$ limit.  Analyses of \Ds\ in bulk samples (and of other observables as well~\cite{carballeira,A13,mgb}) suggest $\cc\sim0.4-1$, that corresponds  to   effects of the cutoff correction basically negligible for $\varepsilon\,\lsim\,0.1$ (\ie, for~$T-\Tc\;\lsim$~8 K if $\Tc\sim$~80 K) but that begin to be appreciable for larger distances to the transition; a value of $\cc\sim0.6$ is also suggested by BCS-like arguments~\cite{carballeira,mgb}. (Our comparisons with data of few-bilayers HTSC  in the next section are also compatible with that strength of the cutoff $\cc\gg0.1$, see later.)

Let us  write the explicit results   obtained by introducing Equations~(\ref{EspectrosInicio}) to~(\ref{EspectrosFinal}) into~(\ref{eq:quantity}) for each  case \Nuctext\,= 1 to 4. The~equations are again long; to shorten them, we found it useful to introduce two auxiliary polynomials $P$ and $Q$  such that:
\begin{equation}
\DsGGL  =   \frac{\electron^2}{32 \hbar \dd \Nuc}
\left[
\frac{P(\varepsilon)}{Q(\varepsilon)}-\frac{P(\cc)}{Q(\cc)}
\right].
\label{eq:DsGGLresultado}
\end{equation}
({The} results without a cutoff may be obtained by removing the second fraction from the formula.) The~explicit expressions we found for the polynomials $P$ and $Q$ are:

For \Nuctext\,= 1:
\begin{equation}
P(\varepsilon)  =  \varepsilon + \gamU \;,
\end{equation}
\begin{equation}
Q(\varepsilon)  =  \varepsilon^2 +2\varepsilon\gamU \;.
\end{equation}

For \Nuctext\,= 2:
\begin{eqnarray}
P(\varepsilon) & = & (4\eps^3 + 12\eps^2 \gamU + 8\eps \gamUcuad) + \nonumber\\ &&
(6\eps^2  +   12\eps \gamU  + 4 \gamUcuad)\; \gamD \;,
\end{eqnarray}\vspace{-18pt}
\begin{eqnarray}
Q(\varepsilon) & = & (\eps^4 + 4\eps^3 \gamU + 4\eps^2 \gamUcuad) + \nonumber\\ &&
(2\eps^3  +   6\eps^2 \gamU  + 4\eps \gamUcuad)\; \gamD \;.
\end{eqnarray}

For \Nuctext\,= 3:
\begin{eqnarray}
P(\varepsilon) & = & (3\eps^5 + 15\eps^4 \gamU + 24\eps^3 \gamUcuad + 12\eps^2 \gamUcubo) + \nonumber\\ &&
 (  10\eps^4  +  40\eps^3 \gamU  + 48\eps^2 \gamUcuad  + 16\eps \gamUcubo)\; \gamD+ \nonumber\\ &&
( 8\eps^3  + 24\eps^2 \gamU  +  19\eps \gamUcuad  + 3 \gamUcubo)\; \gamDcuad \;,
\end{eqnarray}
\begin{eqnarray}
Q(\varepsilon) & = & (\eps^6 + 6\eps^5 \gamU + 12\eps^4 \gamUcuad + 8\eps^3 \gamUcubo)  +\nonumber\\ &&
 ( 4\eps^5  + 20\eps^4 \gamU  +  32\eps^3 \gamUcuad  +   16\eps^2 \gamUcubo)\; \gamD  +\nonumber\\ &&
 ( 4\eps^4  + 16\eps^3 \gamU) \gamDcuad +  ( 19\eps^2 \gamUcuad  + 6\eps \gamUcubo)\; \gamDcuad \;.
\end{eqnarray}

For \Nuctext\,= 4:
\begin{eqnarray}
P(\varepsilon) & = & ( 8\eps^7 + 56\eps^6 \gamU + 144\eps^5 \gamUcuad + 160\eps^4 \gamUcubo  +   64\eps^3 \gamUcuar) + \nonumber\\  &&
( 42\eps^6  +    252\eps^5 \gamU  + 540\eps^4 \gamUcuad  + 480\eps^3 \gamUcubo  +   144\eps^2 \gamUcuar)\; \gamD +\nonumber \\ &&
( 72\eps^5  +    360\eps^4 \gamU  + 616\eps^3 \gamUcuad  + 408\eps^2 \gamUcubo  +   80\eps \gamUcuar )\; \gamDcuad + \nonumber\\ &&
 ( 40\eps^4  + 160\eps^3 \gamU  +    204\eps^2 \gamUcuad  +  88\eps \gamUcubo  + 8 \gamUcuar)\; \gamDcubo \;,
\end{eqnarray}\vspace{-18pt}
\begin{eqnarray}
Q(\varepsilon) & = & (\eps^8 + 8\eps^7 \gamU + 24\eps^6 \gamUcuad + 32\eps^5 \gamUcubo +   16\eps^4 \gamUcuar) +\nonumber\\ &&
 ( 6\eps^7  +   42\eps^6 \gamU  +   108\eps^5 \gamUcuad  + 120\eps^4 \gamUcubo  +   48\eps^3 \gamUcuar)\; \gamD + \nonumber\\ &&
(12\eps^6  + 72\eps^5 \gamU  +   154\eps^4 \gamUcuad  + 136\eps^3 \gamUcubo  +   40\eps^2 \gamUcuar)\;  \gamDcuad + \nonumber\\ &&
( 8\eps^5  + 40\eps^4 \gamU  +    68\eps^3 \gamUcuad  + 44\eps^2 \gamUcubo  +   8\eps \gamUcuar)\; \gamDcubo \;.
\end{eqnarray}


\subsection{Crossover from the Gaussian--Ginzburg--Landau GGL Regime to the Kosterlitz--Thouless KT-like Regime: Ginzburg Number for Few-Bilayers~Superconductors\label{ssec:Gi}}

Up to now, we have considered the GGL approach for the fluctuations. This is  perturbative on the order parameter $\psi$ and thus is only valid for weak fluctuations. However, for~temperatures sufficiently close to the transition, the divergence of fluctuations makes necessary full-critical treatments, which are nonperturbative in $\psi$ \cite{loram,eG}. This is specially important in systems close to 2D, because~ the KT renormalization broadens the effective transition down to the vortex--antivortex binding temperature, \Tkt, thus extending the size of the  full-critical region~\cite{berezinskii1,berezinskii2,KT,HN,R22}.  The temperature for the crossover between the GGL regime and the full-critical one  is usually estimated by the so-called Levanyuk--Ginzburg criterion, \ie, by~calculating the temperature where $|\psi|^4$ contributions to the GL expansion begin to dominate  the $|\psi|^2$ ones, signaling the start of the failure of the perturbation approach~\cite{loram,eG}. This happens at the reduced temperature (usually called Ginzburg number \Gi) at which  the fluctuation specific heat $c_{fl}$ equates the mean field jump of the specific heat at the transition $c_{\rm jump}$ \cite{loram,eG}.  For our present purposes, it is convenient to express this in terms of the GGL paraconductivity (that is in fact proportional to $c_{fl}$  in the GGL approach~\cite{LD,R25}) as $\DsGGL({\scriptsize\varepsilon = \Gi})\, = \,(\pi{\rm e}^2\xi_{ab}^2(0)/4\hbar k_{\rm B})\,c_{\rm jump}$. When our results for \DsGGL\ in few-bilayers HTS are introduced in this condition, we {obtain:}
\begin{equation}
\frac{P(\Gi)}{\Nuc Q(\Gi)} =    \frac{8\pi d\,\xi_{ab}^2(0)\,c_{\rm jump}}{k_{\rm B}}.
\label{cjump}
\end{equation}
{{For}} simplicity, we used in  this equation $\cc\rightarrow\infty$, as~the effect of this parameter is expected to be negligible close to the transition. The~$P$ and $Q$ polynomial functions for each of the \Nuctext\ values are the same as defined in the previous subsection. Note  that $c_{\rm jump}$ is not expected to depend on \Nuctext\ in our functional, and~therefore, these polynomials determine the dependence on \Nuctext\ of \Gi. The~equation itself is implicit, but~it is very easy to  solve it numerically with current~computers.

Let us note already  here that this dependence of \Gi\ with  \Nuctext\ will be an important ingredient in our account of the experimental data in few-bilayers YBCO  in Section~\ref{sec:experimental}, both because it affects  the quality of the fits and because it will allow us to explain  the seemingly  anomalous  dependence with \Nuctext\ of the critical amplitudes of the paraconductivity in the KT-like region, which has been to~our knowledge unexplained until now (see later).

\mbox{}  

\subsection{Kosterlitz--Thouless Paraconductivity \DsKT \label{ssec:DsKT}}

Closer to the transition than $\varepsilon = \Gi$, the fluctuations are expected to be full-critical and dominated by the KT vortex--antivortex correlations and corresponding shift of the critical divergences from \Tc\ down to a new KT transition temperature \Tkt. 

A summary of different attempts to extend the KT theory to infinite-layers superconductors is given, \eg, by~Fischer in~\cite{R21}  (note that the KT theory was originally  formulated for purely 2D systems; no equivalent efforts exist, to~our knowledge, to~extend it for the finite-layers case). As~shown in {Ref.}~\cite{R21}, different authors have proposed routes of extension leading to somewhat different renormalization results, but~quite ample consensus exists in that the relevant temperature dependence of the superconducting coherence length remains as in 2D:
\begin{equation}
\xi_{ab\,\rm KT}(T)\proptosim\exp\sqrt{\frac{b_0(\Tc-\Tkt)}{T-\Tkt}}.
\label{eq:xiKTprop}
\end{equation}
(except when the number of strongly coupled layers forms a set of thickness larger than the vortex coherence length, which is a~limit not relevant to our few-layers case) \cite{R21}. In this expression, $b_0$ is a positive constant, and~the proportionality constant is to be determined from continuity with the GGL regime~\cite{R21,HN}.  It will be convenient for us to re-express Equation~(\ref{eq:xiKTprop}) by stating that the usual reduced temperature $\varepsilon = \ln(T/\Tc)$ has to be replaced by a new expression:
\begin{equation}
\varepsilon_{\rm KT}  =  \varepsilon_{\rm KT}(0) / \exp\sqrt{\frac{4b_0(\Tc-\Tkt)}{T-\Tkt}}.
\end{equation}
where  the proportionality constant needed for continuity of the coherence length at \linebreak$\varepsilon = \varepsilon_{\rm KT} = \Gi$ is:
\begin{equation}
\varepsilon_{\rm KT}(0)  =  \Gi \; \exp\sqrt{\frac{4b_0(\Tc-\Tkt)}{\Tc\exp(\Gi)-\Tkt}}.
\label{epsilonKTcero}
\end{equation}
 {Note} that with these expressions it is now also $\xi_{ab\,\rm KT}(T) = \xi_{ab}(0)\varepsilon_{\rm KT}^{-1/2}$.

The paraconductivity  \DsKT\ in the KT regime of a purely 2D system (\ie, one single layer) has been calculated by, \eg, Halperin and Nelson (HN) in~\cite{HN}. Their proposed equation is the well-known expression $\DsKT = A_{\rm KT}\exp\sqrt{4b_0(\Tc-\Tkt)/(T-\Tkt)}$, which is used in numerous fits to very thin films of cuprates in the tail of the transition (see our Introduction) taking $A_{\rm KT}$ and $4b_0(\Tc-\Tkt)$ as fitting parameters (and sometimes also \Tkt). As~pointed out by HN~\cite{HN}, a different view of their result for \DsKT\ is that  the GGL expression may be used in the KT regime, but~only once the KT temperature dependence for the coherence length is substituted in it. We will apply the same rule to our finite-layered case to~write:
\begin{equation}
\DsKT  =   \frac{e^2}{32 \hbar \dd \Nuc}
\frac{P(\varepsilon_{\rm KT})}{Q(\varepsilon_{\rm KT})},
\end{equation}
which is in correspondence with our Equation~(\ref{eq:DsGGLresultado}) in the limit $\cc\rightarrow\infty$ (that may be used in the KT regime for simplicity and because the effects of \cc\ are expected to be negligible so close to the transition).

It is relevant to mention here that our proposed equation no longer contains a free amplitude parameter $A_{\rm KT}$ as often employed when fitting the classical 2D, one-layer result. This freedom has been removed by the consistency condition of continuity with the GGL fluctuations (in other words, by~the constraint of Equation~(\ref{epsilonKTcero})).


\subsection{Effective Medium Approximation for the Effects of \Tc\ Inhomogenities\label{ssec:EMA}}

When analyzing real experimental data of the critical effects in HTSC (as we do in the next section),   it is mandatory to explore whether the effects of critical temperature inhomogeneities may be affecting the data. This is mainly because of the non-stoichiometric character of HTSC together with the fact that their critical temperatures change with the composition (and the corresponding carrier density) \cite{Tcp}. As any non-stoichiometric  compound may have local random variations of composition, also random local inhomogeneities of \Tc\ may be suspected. It is customary~\cite{usodeEMA1,usodeEMA2,pomar} to take into account the possible effects of the random inhomogeneities  by using the effective medium approximation (EMA)~\cite{MazaEMA}, which for the convenience of the reader, we summarize here. The~EMA gives the \Ds\ of the inhomogeneous system as an implicit equation to be solved numerically \cite{MazaEMA}:
\begin{equation}
\label{eq:inhomogenities}
\int_{-\infty}^{\infty} \;
\frac{\Dspure(\Tc+\tau)-\Ds}
{\Dspure(\Tc+\tau)+2\Ds} 
\exp \left[ \frac{-\tau^2}{ (\delta\Tc/\sqrt{\ln2})^2} \right] \frac{\dd \tau}{\delta \Tc}  =  0,
\end{equation}
where $\Dspure(\Tc+\tau)$ is the paraconductivity of a homogeneous system (\ie, the~equations described in the previous subsection) but calculated with a mean-field critical temperature equal to $\Tc+\tau$ (and KT temperature $\Tkt+\tau$) ({{we take here the quantity} $\Tc-\Tkt$ constant, so that \Tkt\ inhomogeneities mimic the ones in \Tc; we found  this to be sufficient to explain the data without~the need of transforming Equation~(\ref{eq:inhomogenities}) into a double integration.}) Note that this  integration variable $\tau$ runs in Equation~(\ref{eq:inhomogenities}) as a Gaussian random distribution of critical temperature deviations with half-width at half-maximum $\delta\Tc$. The~equation also assumes a 2D geometry. As~is well known,~{Ref}. \cite{MazaEMA} shows that the main effect of the EMA averaging is basically to smooth the predictions of the resistive transition in a vicinity of size $\sim$$\delta\Tc$ around the transition \Tc. Outside of that  window (usually small, see below), the effects are quite~negligible.



\section{Analysis of Experimental~Data\label{sec:experimental}}


In order to compare with some experimental data  our  proposals of a possible theoretical  scenario for the fluctuations in few-bilayers HTSC, we have chosen the pioneering data of  Cieplak \etal.~\cite{cieplak} obtained in the paradigmatic HTSC compound \ybco\ (YBCO).  Cieplak \etal.'s films consist of \Nuctext\ unit cell films of YBCO sandwiched into nonsuperconducting material (many-layers \pbco\  bottom support and top cover), with~samples from the \Nuctext\,= 1 up to \Nuctext\,= 4 cases. We found that~\cite{cieplak} reports in a particularly explicit way the experimental data needed for our~comparisons. 

Another advantage of the data by Cieplak \etal.  in relation to our analyses is that the   background ($\rho_n$) subtraction is one of  the most unambiguous among the reported measurements in few-bilayers HTSC. This is because they explicitly measure the \pbco\ contributions and subtract them from the \ybco\ subsystem, and~because  the latter happens to present well above the transition a linear-in-$T$ behavior of the resistivity~\cite{cieplak}. This allows a quite reliable $\rho_n$ determination (by just doing a linear fit to the data above 1.5$T_{\rm inflect}$, where $T_{\rm inflect}$ is inflection temperature at the transition, \ie, the~one at which $\dd\rho/\dd T$ is maximum).

Before doing a full comparison of our equations with these data, we performed first the common step of fitting the very lower tail of the $\rho(T)\rightarrow0$ limit (that is known to follow the KT-type theories quite well) just using~\cite{HN,cieplak}
\begin{equation}
\label{eq:deriv-KT}
\left( \frac{\dd \ln \rho}{\dd T} \right)^{-2/3}  =  \frac{T-\Tkt}{\sqrt[3]{b_0(\Tc-\Tkt)}}.
\end{equation} 
{The} right-hand side of this equation is the result given by the classical, one-plane KT equation by Halperin and Nelson~\cite{HN}. Its main advantage is that it produces a simple linear fit, which is very unambiguous in its estimate of  the two constants \Tkt\ and $b_0(\Tc-\Tkt)$. Importantly,  this fit leads~\cite{R35,cieplak} to a \Tkt\ value in excellent phenomenological agreement with the appearance  of the non-ohmic voltage--current behavior $V\propto I^3$ expected to occur at the KT transition. We take this value, therefore, as~a solid constraint for the  \Tkt\ to be used in our comparisons (it also produces a value of $b_0(\Tc-\Tkt)$ that we used as first estimate for further refinements in our more complete fits).

We also impose  other physical constraints to our parameter values when comparing our equations to the data: in addition to fixing the mentioned \Tkt, we  impose that  the values obtained for the Ginzburg number  $\Gi$ for each \Nuctext\ are consistent among them; this is equivalent (see Equation~(\ref{cjump})) to requiring the same value of $\xi_{ab}^2(0)c_{\rm jump}$ for all the samples. We also require the values of $b_0$, \gamU, \gamD\ and \cc\ to not vary more than a factor of two from sample to sample. In addition, we require the values of \gamU\ and \gamD\ in each sample to be compatible with the estimates~\cite{R25,A11,A19} of the c-direction GL coherence length amplitude $\xi_c(0) = 1.0$ \AA \,$\pm$ 20\% available for bulk \ybco\ from fluctuation measurements (for equations relating \gamU\ and \gamD\ with $\xi_c(0)$ in bulk HTSC, see~\cite{klemm} or~\cite{R25}). Finally, we tried to use for $\delta\Tc$ the minimum value compatible with the data near the temperature \Tc, \ie, we tried to consider the smallest amount of inhomogeneities reasonably compatible with the data (to make the effects of the finiteness of the layered structure more visible, and {even though increasing} somewhat $\delta\Tc$ could  nominally improve for some samples the quality of the fit near \Tc).

\begin{figure}[h!]
\includegraphics[width =\textwidth]{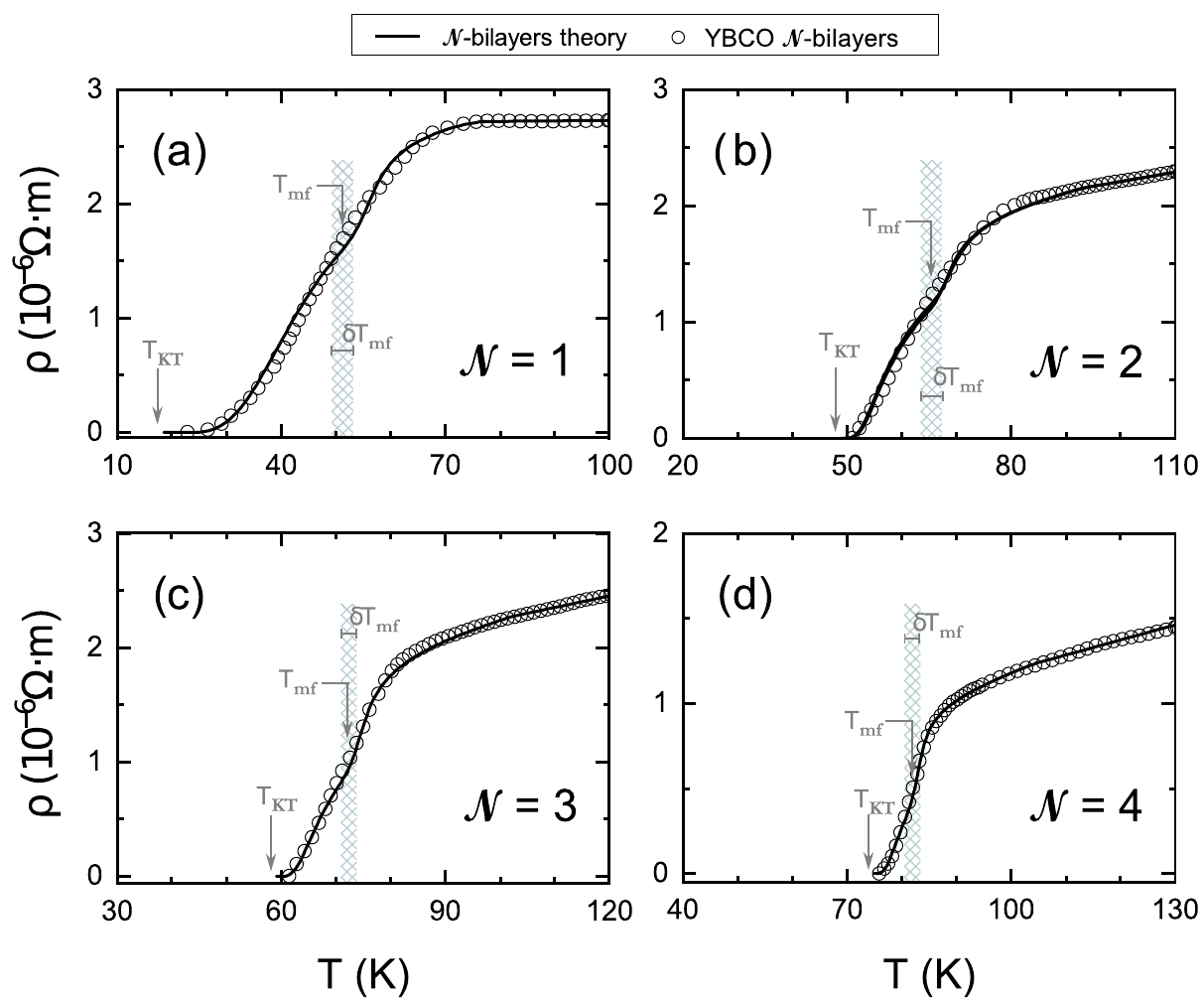}
\caption{{Same data as in}  Figure~\ref{fig:intro} {for the electrical resistivity of samples with {\bf (a)} \Nuctext=1,  {\bf (b)} \Nuctext=2,  {\bf (c)} \Nuctext=3 and  {\bf (d)} \Nuctext=4}  unit cells of superconducting bilayers of \ybco~\cite{cieplak} (open circles) and fits to them using the equations proposed in the present paper for the critical fluctuations in few-bilayers HTSC (solid lines). The~enlargement of the temperature  region in which there is good agreement between data and theory  is evident with respect to the previous approaches shown in Figure~\ref{fig:intro},  mainly above the mean-field critical temperature \Tc. The~shadowed bands correspond to the temperature regions from $\Tc-\delta\Tc$ up to $\Tc+\delta\Tc$, \ie, the~ones affected by the EMA-averaging of \Tc-inhomogeneities. The~employed equations are described in Section~\ref{sec:teoria}. The~general procedures for the fits, and~the discussion of the results, are presented in Section~\ref{sec:experimental}. The~numerical values of the parameters used in these comparisons are listed in Table~\ref{table:uno}.
\label{fig:fitted_R}
}
\end{figure}

\begin{figure}[h!]
\includegraphics[width =.99\textwidth]{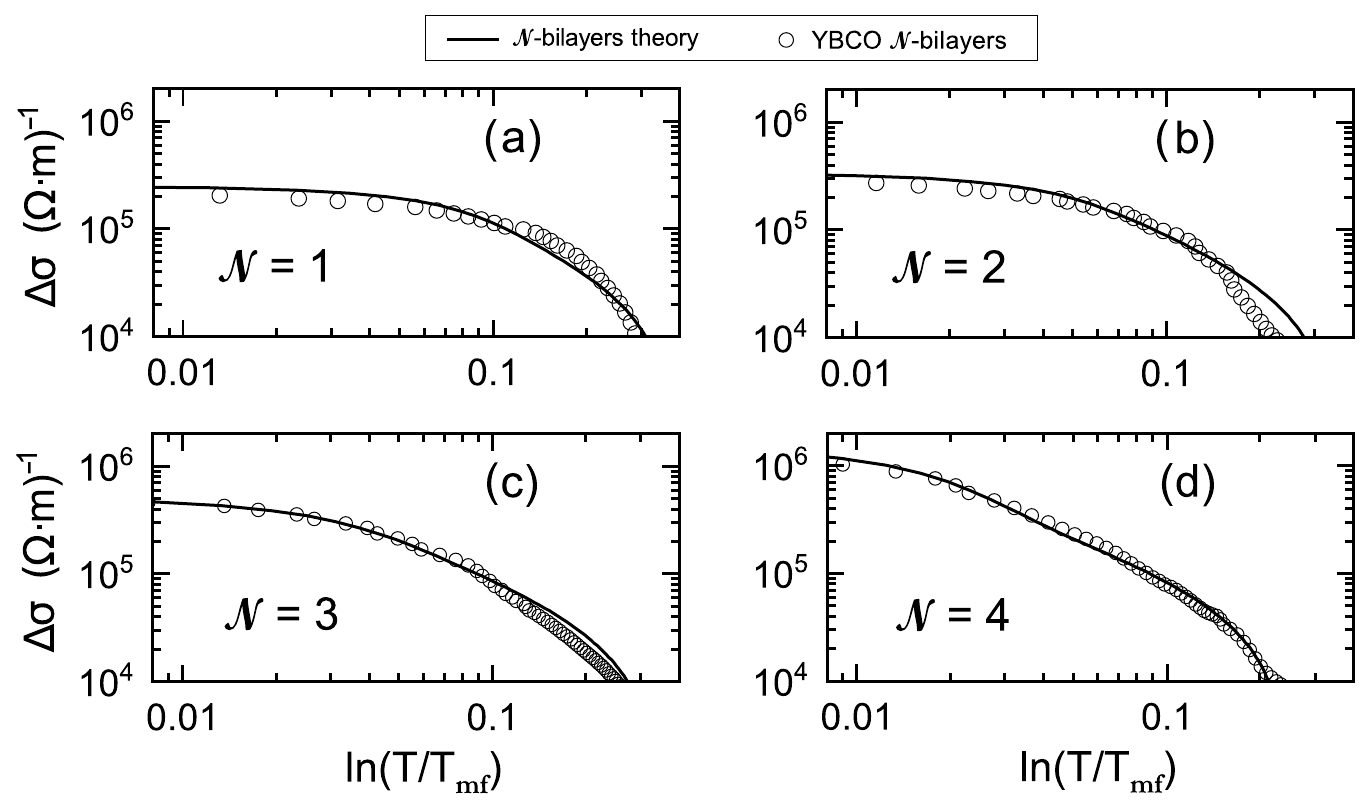}
\caption{{Paraconductivity}  \Ds\ versus reduced temperature $\varepsilon = \ln(T/\Tc)$ corresponding to  the same data~\cite{cieplak} as in Figure~\ref{fig:fitted_R} (open circles) for samples with {\bf (a)} \Nuctext=1,  {\bf (b)} \Nuctext=2,  {\bf (c)} \Nuctext=3 and  {\bf (d)} \Nuctext=4   unit cells of superconducting bilayers of \ybco\     and also the same theory predictions and parameter values as in that Figure (solid lines). This representation is the more usual one in the literature when studying \Ds\ above the transition. Our proposed scenario for the critical fluctuations in few-bilayers HTSC are  in good agreement with the experimental data also in this representation.
\label{fig:fitted_para}}
\end{figure}

The results of our comparisons are shown in Figures~\ref{fig:fitted_R} and~\ref{fig:fitted_para}, and the resulting parameter values are given in Table~\ref{table:uno}. Figure~\ref{fig:fitted_R} illustrates the overall excellent agreement obtained with the $\rho(T)$ transition curves for~all the studied \Nuctext\ cases. This agreement includes not only the KT-like region of the fluctuations but, importantly, also the GGL region (upper part of the transition). In~Figure~\ref{fig:fitted_para}, we draw  the representation that is more usual in the literature when studying \Ds\ above the transition (\Ds\ vs. $\varepsilon$ in log-log axes); it may be seen that the agreement is also excellent in this sensitive~representation.

It is also notable that the good agreement in the KT-like region is achieved although our formulae do not include a free amplitude parameter for \DsKT\ (as already mentioned, in~our approach, the consistency condition of the GGL and KT expressions reduces this degree of freedom in the analysis; in particular, the~dependence of the Ginzburg number \Gi\ on \Nuctext\ is the main factor determining the amplitude of \DsKT). To~further explore this aspect, let us define an ``effective one-layer KT amplitude'', $A_{\rm KT eff}$, as~the amplitude necessary in the \DsKT\ equations of one-layer superconductors to reproduce our few-bilayers results at  a given reference point, that we take as $\varepsilon = \Gi$. Our results in Figures~\ref{fig:fitted_R} and~\ref{fig:fitted_para} and Table~\ref{table:uno} lead, for~\Nuctext\,= 1 to 4, respectively,  to~$A_{\rm KT eff}\simeq 800, 5500, 7000, 8000  \,(\Omega {\rm m})^{-1}$. These numbers are within  about 15\% the ones from the fits using the one-plane KT equation with a totally free amplitude parameter (see our Introduction and the caption of Figure~\ref{fig:intro}). This also includes the one-order-of-magnitude difference between \Nuctext\,= 1 and \Nuctext\,= 4, and~it confirms both  the plausibleness of our proposed scenario and  its usefulness to better understand the KT-like~region.\vspace{-9pt}

In Table~\ref{table:uno}, it may be observed that the Ginzburg number \Gi\ increases as \Nuctext\ decreases (so that the largest full-critical region above \Tc\ is the one of \Nuctext\,= 1, as~it also happens with the full-critical region below \Tc). Note also that \Gi\ is for \Nuctext\,= 4 already close to the $\sim$0.01 value usually found for bulk YBCO near optimal doping~\cite{R25,eG}.

We  found the fit to be quite sensitive to the value of the ratio $\gamU/\gamD$, which converges for all \Nuctext\ to a value $\simeq30$. In~bulk, infinite-layers YBCO samples, the analyses of \Ds\ do not really distinguish much~\cite{R25} between  $\gamU/\gamD = 1$ and 30 or even  $\sim$100 (as the differences are within the experimental uncertainties in \Ds), but~ the present analyses of the few-bilayers samples seem to open a way for a more strict determination  for that~ratio.

The shadowed bands in Figure~\ref{fig:fitted_R} are the temperature regions from $\Tc-\delta\Tc$ up to $\Tc+\delta\Tc$, \ie, the~ones affected by the EMA-averaging of \Tc-inhomogeneities. Let us note that increasing somewhat the employed dispersions $\delta\Tc$ could improve the agreement with the data around these bands. However, this would make less visible the effects of our few-layers considerations, which are the main focus of this paper. Note that our $\delta\Tc$ values in Table~\ref{table:uno} are  comparable with the ones usually found in the best bulk YBCO samples near optimal doping, \ie, the~theory does not require anomalously large $\delta\Tc$ values to explain the data,  even for \Nuctext\,= 1  (contrarily to the statement in this regard by Cieplak \etal.~\cite{cieplak} in their early work, which is caused by their use of the infinite-layers theory).

In an Appendix we briefly summarize a similar comparison of our equations with available data~\cite{bscco_r1} in few-bilayers  films of the HTSC compund \bscco\ (BSCCO).





\section{{Conclusions:  Some Implications and Open~Aspects}\label{sec:conclusiones}}


To sum up, we have studied the critical fluctuations near the resistive transition  of very thin films of high-temperature cuprate superconductors composed of   a number \Nuctext\ of only a few unit cells of superconducting bilayers. For~that, we explicitly solved the fluctuation spectrum of  a Gaussian--Ginzburg--Landau model for few-bilayers superconductors, considering two alternating Josephson interlayer interaction strengths. We then obtained the corresponding explicit expressions for the paraconductivity \Ds\ above the mean-field transition temperature, \Tc, for~various values of \Nuctext. We also obtained expresions, within~the same modelization, for~ the crossover from the Gaussian regime to the Kosterlitz--Thouless--type full-critical regime of the fluctuations by~calculating the Ginzburg number \Gi\  and its dependence on \Nuctext. We also proposed  expressions for \Ds\ in the KT-like regime that are coherent with that~crossover. 

We then compared these theory results with available data in \ybco\ few-bilayers systems with \Nuctext\,= 1 to 4, for~which we have used the paradigmatic data reported  by Cieplak \etal\ in Ref.~\cite{cieplak}. That comparison leads to a good agreement that extends over a significantly larger temperature region than previous theory scenarios  based on either one-layer or infinite-layers models. It also justifies the seemingly  anomalous critical amplitudes of the paraconductivity in the KT-like region.  {Available data in few-bilayers \bscco~\cite{bscco_r1}  also display  agreement with our proposed equations.}

{
In addition to their interest in understanding the critical phenomena in few-bilayers HTSC, these results may be also useful to better understand some general characteristics of the  pairing in these superconductors. For~instance, they suggest that the locus of the superconducting wavefunction is each \cuodos\ individual plane (rather than structureless biplanes) in line with the considerations about the role of interlayer interactions in the pairing energy balances by, \eg, Refs.~\cite{leggett,LeggettMIR,PhilAnderson,Urbana}. They also support the relevance of the phase fluctuations  in the  tail of the transition and~thus its influence on the verification of the Uemura plot in HTSC~\cite{Uemura,N12,N3,N16,A13,B1}, while above the transition, both  phase and amplitude fluctuations\linebreak coexist~\cite{N12,N3,B1,leggett,Urbana}.
}

{
In addition, our results suggest some aspects that would merit additional research in the future. For~instance, we could study in few-bilayer HTSC the  fluctuation effects in  other properties, such as in magnetoresistivity,  magnetic susceptibility or the specific heat. The~two later would present the useful theory advantage of its fluctuation roundings being basically proportional to  \Ds\ in the GGL regime~\cite{klemm,R25,alberto}, but to our knowledge, the fluctuation effects in them have not been measured up to now  in very thin films above \Tc\ because of the smallness of the samples (for the associated experimental problems, see, \eg, Ref.~\cite{cotonchi}). }

{
It would be also interesting to extend these studies to Fe-based superconductors. They are also layered and present a broad range of anisotropies and interlayer--interaction strengths. Note that few-layer films have been already created for at least the \linebreak122 pnictide~\cite{122-1} and FeSe~\cite{FeSe-1} families, and~they should be possible for the 1111 pnictide family~\cite{1111-films-1,1111-films-2} (for a review, see~\cite{1111-films-0}; also, single-crystals of the 1111 family could be viewed as heterostructures at the atomic limit~\cite{1111-HsAL}). We also emphasize that for studying  few-layer Fe-based superconductors, it would be crucial to extend our present calculations  with multiband effects by~considering multicomponent intercoupled order parameters~\cite{multibands-1,multibands-2}. In addition, in~some cases, interface superconductivity states may be important, as~in the outer layers of the Fe(Se,Te)-type superconductors~\cite{interfaceSC-1,interfaceSC-2}.
}

\vspace{6pt}

{\bf Author contributions: } {Conceptualization, M.V.R.; Investigation, M.M.B. and M.V.R.; Writing---original draft,  M.M.B. and M.V.R.  All authors have read and agreed to the published version of the~manuscript.}

{\bf Acknowledgements:} {This work was supported by the Agencia Estatal de Investigación (AEI) and Fondo Europeo de Desarrollo Regional (FEDER) through project PID2019-104296GB-100, by~Xunta de Galicia (grant GRC number ED431C 2018/11) and iMATUS (2021 internal project RL3). MMB was supported by Ministerio de Universidades of Spain through the National Program FPU (Grant Number FPU19/05266).}

\clearpage

{\bf Abbreviations:} {
The following abbreviations are used in this manuscript:

\noindent 
\begin{tabular}{@{}ll}
2D & two-dimensional\\
BCS & Bardeen--Cooper--Schieffer\\
GGL & Gaussian--Ginzburg--Landau\\
GL & Ginzburg--Landau\\
HN & Halperin--Nelson\\
HTSC & high-temperature {superconducting cuprates}\\
KT &  Kosterlitz--Thouless\\
LD & Lawrence--Doniach\\
PBCO & \pbco\\
YBCO & \ybco\\
BSCCO & \bscco\\
\end{tabular}
}

\mbox{}

\clearpage

\clearpage

\clearpage

\clearpage

\begin{table}[h!]
\caption{{Parameter}  values resulting from the fits represented in Figures~\ref{fig:fitted_R} and ~\ref{fig:fitted_para}. Note that $\gamD$ does not appear in the equations for \Nuctext\,= 1.  \label{table:uno}}
\begin{center}\begin{tabular}{ccccccccc}
\hline
\boldmath\Nuctext & \boldmath \textbf{\Tkt\ (K) }&\boldmath\textbf{  \Tc\ (K) }  & \boldmath\textbf{$\delta\Tc\ (K)$ }& \textbf{Gi} & \boldmath$b_0$ & \boldmath\gamU & \boldmath\gamU/\gamD &  \boldmath\cc  \\ \hline
  1   &     17.5 &     51.2  &   2.5  &   0.065    &  7.8     & 0.55       & ---  &  0.40   \\
  2  &     46.9 &     65.4  &     2 &       0.035    &  4.1     & 0.45    & 30 &  0.35  \\
  3 &     58.1  &     71.8    &   2      &   0.02   &   4.5        &  0.30        & 30 &  0.35 \\
  4  &     74.3 &     82.2    &   1.5  &   0.01    &  5.6    &  0.60       & 30     &  0.25   \\
\hline
\end{tabular}\end{center}
\end{table}

\clearpage

\section*{Appendix: 
Comparison with Few-Bilayer\\ \boldmath\bscco\ Ultra-Thin Films
}


{     
It may be interesting to check our model also in other  HTSC with different compositions and anisotropies for which few-bilayer films have been produced such as, \eg,  \bscco\ (BSCCO) \cite{bscco_s1,bscco_s2,bscco_r3,bscco_r1,bscco_r2}. The \cuodos\ planes in BSCCO may be  considered to form a bilayered-like structure, with~an average interlayer distance $d = 7.7$ \AA. BSCCO is known to be more anisotropic than YBCO, to~the point that  $\gamU, \gamD\simeq0$ may be suspected to be a good approximation~\cite{R4,R35,R7,R3,cieplak,bscco_s1,bscco_s2,bscco_r3,bscco_r1,bscco_r2}. In this limit, as~it could be expected, the application of our equations for the GGL regime simply leads to $\DsGGL({\scriptsize\gamU = \gamD = 0})  =  (\electron^2/8\hbar d)(1/\eps-1/\cc)$. In~other words, \DsGGL\ recovers a pure 2D  exponent, with~appropriate thickness normalization and  total-energy cutoff regularization.
} 

\begin{figure}[h!]
\centering 
\includegraphics[width =\textwidth]{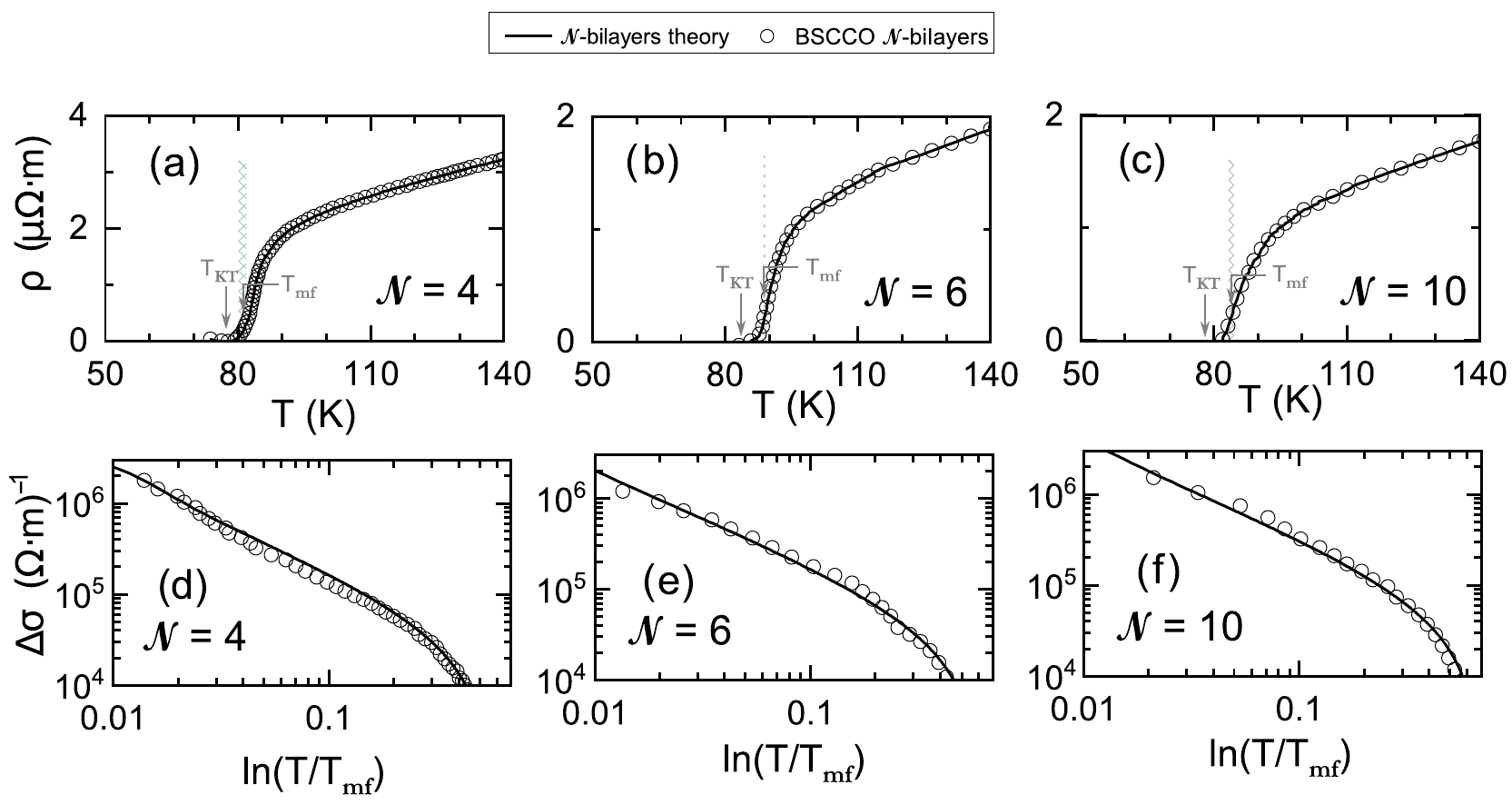}
\caption{
{
{Resistivity}  $\rho$ versus temperature $T$ (open circles, panels (\textbf{a}--\textbf{c})) and paraconductivity \Ds\ versus  $\eps = \ln(T/\Tc)$ (open circles, panels (\textbf{d}--\textbf{f}))  obtained experimentally by Zhao \etal.~\cite{bscco_r1} in ultra-thin films of \bscco, for~three representative number of \cuodos\ bilayers \Nuctext. The~solid lines are fits to these data using our equations. We used as parameters, for~$\Nuctext\,= 4, 6, 10,$ respectively, the~following:  $\Tc = 81.2, 88.8, 83.9$~K;  $\Tkt = 77.3, 83.5, 78.0$~K; $\delta\Tc = 0.7, 0.25, 0.65$~K; $b_0 = 4, 6, 5$; $\cc = 0.55, 0.6, 0.68$; $\Gi = 0.009, 0.005, 0.004$. The~normal-state background $\rho_n$ was obtained by linear extrapolation of the data above 160~K (we observe  no significant changes when varying this  temperature). The shadowed bands correspond to the temperature regions from $\Tc-\delta\Tc$ up to $\Tc+\delta\Tc$.
}
\label{fig:appendix}}
\end{figure}

{     
We confirmed that  our equations, taken with $\gamU, \gamD = 0$,  do agree with experiments in few-bilayer BSCCO. For~that, we used the measurements of $\rho(T)$  obtained by  Zhao \etal.~\cite{bscco_r1} in very high-quality ultra-thin films (with \Nuctext\,= 4 to 20) of BSCCO. In~Figure~\ref{fig:appendix}, we show such comparisons for some representative values of \Nuctext.  The~obtained parameter values are given in the figure caption.  We conclude that this comparison  again supports  the plausibility of our proposed scenario.
} 

\end{document}